\documentclass[12pt]{article}

\begin{document}

%******************************************

%*******************************
%*******************************

\begin{flushright}
MIT-CTP-3378 \\
YITP-SB-03-21\\\
LA-UR-03-2054\\
hep-ph/0305241
\end{flushright}

% \vspace{0.25in}

\begin{center}
{\Large \bf Problems of the rotating-torsion-balance limit on the photon mass}
\end{center}

\vspace{.165in}

{\baselineskip=16pt \centerline{\large Alfred Scharff Goldhaber$^*$\footnote{
goldhab@insti.physics.sunysb.edu, 
on leave from C.N. Yang Institute for Theoretical Physics, State
University of New York, 
Stony Brook, NY 11794-3840} and Michael Martin
Nieto$^\dagger$\footnote{mmn@lanl.gov}} 
\bigskip
\centerline{\it $^*$Center for Theoretical Physics}
\centerline{\it Massachusetts Institute of Technology} \centerline{\it
Cambridge, MA 02139}
\smallskip
\centerline{\it and}
\smallskip
\centerline{\it $^\dagger$Theoretical Division (T-8, MS-B285)}
\centerline{\it University of California}
\centerline{\it Los Alamos National Laboratory} \centerline{\it
Los Alamos, NM 87545}}

\vspace{.33in}

\vspace{.165in}

%****************************************
\begin{abstract}
We discuss the problems (and the promise) of the ingenious method
introduced by  Lakes, and recently improved on by Luo, 
to detect a possible small photon mass $\mu$ by measuring the ambient
magnetic vector potential from large scale magnetic fields. We also
point out how an improved ``indirect'' limit can be obtained using modern
measurements of astrophysical magnetic fields and plasmas and that a 
good ``direct'' limit exists using properties of the solar wind. 

\end{abstract}

%****************************************
\newpage

Recently Luo et al.\cite{luo} improved an ingenious method  of  Lakes
\cite{lakes} to detect a possible small photon  mass $\mu$.  In the
Proca ($\mu ={\rm constant}$) formulation, nonzero
$\mu$ fixes the Lorentz gauge for electrodynamics,
and thus makes unique the vector potential {\bf A} at any point due to
specified sources.  The  
$\mu^2A^2$  Proca term in the Lagrangian implies a
torque  on a loop of magnetic flux from the
ambient magnetic vector potential ${\bf A}_{\rm amb}$, analogous to
the torque on a loop of electric current from 
an ambient magnetic field.   The torque 
{\boldmath${\tau}$}$=${\boldmath${\nu\times}$}$\mu^2\mathbf{A}_{\rm amb}$
acts on {\boldmath${\nu}$}, the `vector-potential dipole moment' of the
flux loop.  As one knows {\boldmath${\nu}$},  measuring or
limiting  {\boldmath${\tau}$} yields  
$(\mu^2\mathbf{A}_{\rm amb})$.  Determining 
$\mathbf{A}_{\rm amb}$ then places a value on $\mu$.   

A typical value of $\mathbf{A}_{\rm amb}$ in a given region can be
very large,  $A\sim |\langle {\bf B} \rangle| L$ (where $L$ is a
characteristic size of the region over which ${\bf B}$ is
approximately uniform). Even small $\langle {\bf B} \rangle$ can be
overcome by large enough $L$ to give large $A_{\rm amb}$, hence
low  $\mu$.  

Lakes \cite{lakes} already noted a source of statistical error --  at
any particular location within a large region of approximately uniform
${\bf B}$ (whose exact boundaries are poorly 
specified), one knows neither the  direction nor the  magnitude of 
$\mathbf{A}_{\rm amb}$.  Lakes looked for
diurnal variation in the torque  on his toroidal magnet.  So, for  
$\mathbf{A}$ closely aligned with the rotation axis of the Earth, he 
would have been insensitive to $\mu$.  The new  improvement \cite{luo}
was rotating the axis of the magnet, allowing  detection of all
projections of $\mathbf{A}$, and also 100 times greater sensitivity to the
signal thanks to a new, adjustable rotation frequency 
for lock-in detection. Eliminating the angular ambiguity reduces the
statistical uncertainty in the Lakes method by roughly a factor $\sqrt2$.
Refs. \cite{luo,lakes} do not account for this uncertainty explicitly 
in their quoted limits.

%**********

Though original and potentially promising for the
future, these works \cite{luo,lakes} neither provide the best
available limit on $\mu^2A$ nor a reliable limit at all on $\mu$.

%***********

{\bf (1)}  
For specified sources the Proca equation in vacuum
implies exponential Yukawa damping of the magnetic vector
potential and field on the scale of the reduced photon Compton wavelength
$^-\!\!\!\!\lambda_C=1/\mu$. However, 
in the presence of  plasma a static magnetic field may take exactly
the form it would have in $\mu=0$ magnetohydrodynamics, {\it provided} 
\cite{wp,gn} the plasma supports a current 
$\mathbf{J}$ 
that exactly cancels the  `pseudocurrent' 
$-\mu^2\mathbf{A}/\mu_0$ 
induced by the photon mass. Thus, if we place a limit on plasma
currents everywhere in a region larger than some putative value of
$1/\mu$, we place the same limit on $\mu^2A$. 

%************************

{\bf (2)}  Using the above, we can obtain a stronger limit. 
For the largest available $A$ (coming from a typical $\mathbf{B}$
over the dimensions of clusters like Coma \cite{coma,den}), 
we require a limit on the intergalactic plasma current, obtainable
from the same astrophysical data used in \cite{luo,lakes} to estimate $A$. 
The mean electron density is $\leq 0.01$ cm$^{-3}$ \cite{den}.   The electron
temperature is about 5 keV (higher in places) \cite{den},  yielding a (more
than generous!) velocity bound on the order of  0.1 c.   This allows a 
current density $< 5 \times 10^{-8}$ A/m$^2$, roughly a factor 200 
smaller than the pseudocurrent allowed by the result of Luo et
al. \cite{luo}.  This current density 
limit of course applies everywhere, including all places 
where $A$ has its typical size.  The resulting limit for
$\hbar\mu/c$ is about $10^{-52}$ g,  
or $^-\!\!\!\!\lambda_C > 4\times 10^{9}$ km, almost 30 Astronomical Units.
Uncertainty about the degree of inhomogeneity in Coma or even in 
in our local galactic cluster makes it hard to quote a definite
result, but it is unlikely to be worse than the claim of \cite{luo}. 

%************************

{\bf (3)}
Of course, anywhere the plasma density
becomes unusually small, including any large vacancies in the plasma
allowed by our ignorance about inhomogeneity \cite{den}, the vacuum
exponential decay applies.  If we happened to be in such a vacancy
then $A$ at our location could be arbitrarily small, and hence the lab
limit on $\mu^2 A$ would give no constraint on $\mu$.

%************************

{\bf (4)}  
Although the torque method cannot yet yield a solid limit on $\mu$,
surely the true limit is smaller than that from the best direct
observations, but we have no clear idea by how much.  The
best direct limit we know comes from Ryutov \cite{ryutov}, (who used a
generous upper bound on the $\mu^2A^2$ energy of the solar wind magnetic
field),
$\mu<10^{-49}$ g or $^-\!\!\!\!\lambda_C=3\times !0^6$ km, 
about 5 solar radii \cite{ade}. 

We thank Philipp Kronberg for information on intergalactic fields and
plasma and  Bill Feldman, Peter Gary, and Jack Gosling for 
information on the solar wind.
This work was supported by NSF grant
PHY-0140192 and US DOE grant DF-FC02-94ER40818 (ASG) and also 
by US DOE contract W-7405-ENG-36 (MMN).

\vspace{.165in}

%******************************************
%******************************************

%****************************** 

\end{document}